\documentclass{ws-mpla}
\usepackage[super]{cite}
\usepackage{graphicx}
\begin{document}

\markboth{Kevin J. Ludwick}
{The Viability of Phantom Dark Energy:  A Brief Review}

\catchline{}{}{}{}{}

\title{The Viability of Phantom Dark Energy:  A Review\\
}

\author{\footnotesize Kevin J. Ludwick
}

\address{Department of Chemistry and Physics, LaGrange College, 601 Broad St\\
LaGrange, GA 30240,
USA
kludwick@lagrange.edu}



\maketitle


\begin{abstract}
In this brief review, we examine the theoretical consistency and viability of phantom dark energy.  Almost all data sets from cosmological probes are compatible with dark 
energy of the phantom variety (i.e., equation-of-state parameter $w<-1$) and may even favor evolving dark energy, and since we expect every physical entity to have some kind of field description, we set out
 to examine the case for phantom dark energy as a field theory.  
 We discuss the many attempts at frameworks that may 
mitigate and eliminate theoretical pathologies associated with phantom dark energy.  We also examine frameworks that provide an apparent measurement $w<-1$ while 
avoiding the need for a phantom field theory.  


\end{abstract}

\ccode{04.62.+v, 04.25.Nx, 11.10.Ef, 95.36.+x, 98.80.-k, 04.50.Kd}

\section{Introduction}	

A recent milestone in observational cosmology happened when the High-z Supernova Search Team in 1998 \cite{Riess} and the Supernova Cosmology Project in 1999 \cite{SCP} published observations of the emission spectra of Type Ia supernovae indicating that the universe's rate of expansion is increasing.  
Galaxy surveys and the late-time integrated Sachs-Wolfe effect also give evidence for the universe's acceleration.  Thus, "dark energy" was proposed 
as the pervasive energy in the universe necessary to produce the outward force that causes this acceleration, which has been observationally tested and vetted 
since its discovery.  The 2011 Nobel Prize in Physics was awarded to Schmidt, Riess, and Perlmutter for their pioneering work leading to the discovery of dark energy.  
The present-day equation-of-state parameter $w$ from the equation of state most frequently tested by cosmological probes, $p=w \rho$ with constant $w$, assuming a 
flat universe and a perfect fluid representing dark energy, has been constrained by Planck in early 2015 to be $w=-1.006 \pm 0.045$ \cite{Planck}, and Planck's 2013 
value is $w=-1.13^{+0.13}_{-0.10}$ \cite{Planck2013}.  The value from the Nine-Year Wilkinson Microwave Anisotropy Probe (WMAP9), combining data from 
WMAP, the cosmic microwave background (CMB), baryonic acoustic oscillations (BAO), supernova measurements, and $H_0$ measurements, is 
$w = -1.084 \pm 0.063$ \cite{WMAP9}.  From these reported values, the prospect of $w<-1$ is clearly a distinct possibility, and under other assumptions (such as a spatially curved universe), 
the window reported for $w$ does not always include the value for the cosmological constant (CC) model, $w=-1$.  

Dark energy with $w<-1$ is often called "phantom dark energy."  It indicates that the energy density of dark energy is increasing over time (as opposed to being constant as 
in the CC model).  If the universe really is accelerating due to phantom dark energy, the universe may end in a big rip \cite{Caldwell}, a 
little rip \cite{littlerip}, a pseudo-rip \cite{pseudorip}, and several types of future signularities can occur \cite{hep-th/0501025}.  
In a model with a constant $w<-1$, a big rip will occur, which means 
that the scale factor of the universe $a$ will reach infinity in a finite time from now.  An energy source that continuously causes an increasing acceleration rate seems unphysical, 
especially since it can lead to the ripping apart of space-time itself in this way, which is at least unpalatable in some sense.  More aspects detrimental to physicality are revealed 
when phantom dark energy is examined as a field. 

All physical phenomena are expected to have a microscopic theory with a field description, and phantom dark energy is no exception.  
In the following sections, we will discuss the theoretical viability of phantom dark energy as a field.  We will first outline the conventional approach to adding a scalar field 
for phantom dark energy in the standard cosmological metric, the
flat Friedmann-Lema\^{i}tre-Robertson-Walker (FLRW) metric, and how the phantom field is incompatible with the metric.  We will then outline the methods around this 
obstacle, such as k-essence and scalar-tensor theories.  
We then discuss the pathologies that come along with these attempts to ameliorate these difficulties.  We then discuss attempts with conventional field models and frameworks 
 to give an 
appearance of $w<-1$ as far as cosmological measurements are concerned.  We then summarize which methods lead to viable theories for phantom dark energy. 



\section{Scalar Field Phantom Dark Energy in Flat FLRW Space}

The simplest field is a scalar field.  Consider the Einstein-Hilbert action for general relativity with a complex scalar field ($c=1$):
\begin{equation}
\label{action}
S = \int d^4 x \sqrt{-g} \left[ \frac{R}{16 \pi G} - \frac{1}{2} g^{\mu \nu} \nabla_\mu \phi^* \nabla_\nu \phi - V(|\phi|) \right] + S_m,
\end{equation}
where the first term is the usual contribution to the Einstein tensor, the second and third terms are the contribution to the scalar field dark energy, and $S_m$ is the action 
for the rest of the components of the energy-momentum tensor $T_{\mu \nu}$.  Minimizing the action leads to Einstein's equation, 
\begin{equation}
\label{Einstein}
R_{\mu \nu} - \frac{1}{2} R g_{\mu \nu} = 8\pi G(T_{\mu \nu}[\phi]+T_{\mu \nu}[m]),
\end{equation}
where $T_{\mu \nu}[\phi] = -2 \frac{\delta \mathcal{L}_\phi}{\delta g^{\mu \nu}}+ g_{\mu \nu} \mathcal{L}_\phi$.

Assuming dark energy is spatially homogeneous as a perfect fluid, the density $\rho_\phi$ and pressure $P_\phi$ for the scalar field are  
\begin{equation}
\label{scalarfluid}
\rho_{\phi} = \frac{\dot{|\phi|^2}}{2 a^2}+ V(|\phi|), \quad P_{\phi} = \frac{\dot{|\phi|^2}}{2 a^2} - V(|\phi|).
\end{equation}
We used the flat FLRW metric
\begin{equation}
\label{FLRW}
ds^2= a^2(\tau) \left[- d\tau^2 + dx^i dx_i \right],
\end{equation}
and $\cdot$ represents differentiation with respect to $\tau$.  

The kinetic energy for the scalar field from the Lagrangian density $\mathcal{L}_\phi$ is $ - \frac{1}{2} g^{\mu \nu} \nabla_\mu \phi^* \nabla_\nu \phi = \frac{\dot{|\phi|^2}}{2 a^2}$.  
The equation-of-state parameter $w=\frac{P}{\rho}$ for dark energy is 
\begin{equation}
\label{w}
w_\phi=\frac{\frac{\dot{|\phi|^2}}{2 a^2} - V(|\phi|)}{\frac{\dot{|\phi|^2}}{2 a^2} + V(|\phi|)},
\end{equation}
and one can see that $w_\phi<-1$ and the physically reasonable condition $\rho_\phi \geq 0$ imply $\rho_\phi+P_\phi=\frac{\dot{|\phi|^2}}{a^2} = 2 ~\mathrm{KE}_\phi < 0$, which 
mathematically cannot be true for a complex or real scalar field.  

\section{Wrong-Sign Kinetic Term}

What is usually done to allow for compatibility of the scalar field with flat FLRW space is to flip the sign in front of the kinetic energy term in the Lagrangian density.  Then the 
ratio for $w_\phi$ becomes
\begin{equation}
\label{wrongsignw}
w_\phi=\frac{-\frac{\dot{|\phi|^2}}{2 a^2} - V(|\phi|)}{-\frac{\dot{|\phi|^2}}{2 a^2} + V(|\phi|)},
\end{equation}
and $w_\phi<-1$ is mathematically allowed.  

\section{K-Essence}

The wrong-sign kinetic term approach is a specific instance of k-essence, which in general is the approach that replaces the kinetic term in the Lagrangian density 
with a function of the kinetic term and the field:  $F(X, \phi)$, where $X=-\frac{1}{2} g^{\mu \nu} \partial_{\mu} \phi \partial_{\nu} \phi$ is the kinetic term 
\cite{astro-ph/0006373}.  In this approach, a negative kinetic term is generally required in order to have $w<-1$ \cite{Caldwell}.  A two-field model is used in the quintom approach 
\cite{0404224}, one a canonical field and one a phantom field.

In general, k-essence theories are plagued with caustics in the non-linear regime so that the field is not single-valued and second derivatives of the field are divergent 
\cite{hep-th/0302199, 1706.01706}.  One possible way around this problem is a dynamical metric that provides backreaction that prevents the formation of caustics \cite{1602.00735}, 
and the introduction of a complex scalar field prevents the divergence from developing in real (as opposed to imaginary) time \cite{1704.03367}.  And two exceptions to this 
generic development of caustics are the Born-Infeld theory and Sen's Lagrangian for the tachyon \cite{hep-th/0302199}.  However, k-essence field theories suffer from other 
pathologies, as we discuss below.

\section{Problems with Wrong-Sign Field Theories and Phantom Pathologies}

Ultimately, these field theories with wrong-sign kinetic terms (called "ghost" field theories) are unstable when coupled to matter in any way, 
and the dark energy field must at least be coupled gravitationally.  An infinite decay rate of the vacuum is a consequence of this coupling, regardless of 
whether or not the mass is above a cut-off 
scale \cite{1406.4550}, and this infinite decay rate is clearly not observed, despite some theoretical connection between k-essence and the effective field theory of a superconducting 
membrane \cite{1608.06540}. 
Either the phantom ghost field has positive density and violates unitarity, rendering it unphysical, or unitarity is satisfied and the density is negative, which leads to vacuum 
instability and unbounded decay of the vacuum via the ghost field \cite{CHT}.  

\subsection{Null Energy Condition}

The null energy condition (NEC) is a constraint on the energy-momentum tensor:  $T_{\mu \nu} n^{\mu} n^{\nu} \geq 0$, where $n^{\mu}$ is a null vector.   
Violation of the NEC is often used as an indicator of phantom behavior of a field.  For causal, Lorentz-invariant scalar theories, the NEC is sufficient to 
determine the classical stability of a theory \cite{hep-th/0606091}.  (Effective field theories that violate the NEC while remaining stable must lack isotropy and have 
superluminal modes \cite{hep-th/0512260}.)
Buniy {\it et al} show that for causal, Lorentz invariant theories (minimally or non-minimally coupled 
scalar and gauge field theories with second-order equations of motion), NEC violation implies classical 
instability with respect to the formation of gradients, and violation of the quantum averaged NEC, $\left<\alpha| T_{\mu \nu} n^{\mu} n^{\nu} | \alpha \right> \geq 0$,  involving
 the bare energy-momentum tensor implies local instability of the quantum state \cite{hep-th/0606091}.  
 
 If dark energy is modeled as a perfect fluid, NEC violation implies a complex speed of fluid propagation or a clumping instability of the fluid.  And for an isolated system that is 
 homogeneous and isotropic, violation of the NEC implies a negative temperature, or an entropy which decreases with energy.  This implies negative kinetic energy in order 
 for the partition function of the system to converge \cite{hep-th/0606091}.  Field theories with negative kinetic energy will roll up a potential hill instead of rolling down a potential well.  

\subsection{Various Phantom Field Theories}

Many theories of phantom dark energy which avoid most pathologies are possible (some among them being theories of vector dark 
energy, Dvali-Gabadadze-Porrati (DGP) branes, Dirac-Born-Infeld (DBI), galileon, kinetic braiding, and other scalar-tensor 
varieties \cite{0709.2399, 0405267, 0801.1486, 0805.4229, 0005016, 1008.0048, 1989, 0412320, 1311.5889}).  
They usually feature at least one of either ghosts, superluminal modes, Lorentz violation, non-locality, or instability to quantum corrections.  

It is possible for a k-essence or $F(X, \phi)$ theory to obey unitarity at tree level and to violate the NEC while remaining ghost-free and free of gradient instabilities at 
the expense of a fine-tuned theory 
that necessarily invokes higher-order irrelevant operators and imposes either a shift symmetry or technically unnatural small operator coefficients in the low-energy effective theory 
\cite{1703.00025}.  We note that Lorentz symmetry is not maintained in this model, however.  

$F(X, \phi)$ theories with ghost condensation can avoid superluminal modes while still violating the NEC by the inclusion of higher-derivative spatial gradient terms to stabilize 
the dispersion relation \cite{hep-th/0312099}.  However, there is no Lorentz-invariant vacuum in this theory.  DGP brane and 
kinetic braiding theories are also generally known to violate Lorentz symmetry. 

The conformal galileon theory \cite{0811.2197} violates the NEC while remaining stable against perturbations and quantum corrections.  However, similar to the ghost 
condensate, there is no Lorentz-invariant vacuum. 

The DBI galileon theory \cite{1003.5917} violates the NEC and is stable against radiative corrections, and the $2 \rightarrow 2$ tree-level scattering amplitude
 satisfies analyticity requirements 
for locality \cite{1212.3607}.  However, perturbations of its Poincar\'{e}-invariant vacuum result in superluminal 
perturbation propagation.  

An attempt \cite{1311.5889} to overcome these issues involves a theory that violates the NEC while maintaining a Poincar\'{e}-invariant vacuum with stable, sub-luminal 
perturbations.  However, because the theory breaks dilation-invariance, the authors of this work suspect the theory to be unstable to quantum corrections.  

In Lorentz-invariant theories, vacuum stability demands the positivity of not only the kinetic term but also the mass terms and self-interaction terms \cite{hep-th/0602178}.  
Effective field 
theories of any variety (not restricted to scalar field theories) which are Lorentz-invariant must have kinetic terms and leading interaction terms which are positive in order 
to ensure that fluctuations around translationally invariant backgrounds do not propagate superluminally.  These superluminal propagations do not allow 
for a Lorentz-invariant notion of causality, and such field theories turn out to be non-local and do not meet $S$-matrix analyticity requirements.  Even theories with positive 
kinetic terms but negative higher-order derivative interactions suffer from these pathologies, and they are flawed in general in both the ultraviolet (UV) and infrared (IR) 
scales \cite{hep-th/0602178}.  

For a ghost theory obeying Lorentz invariance, superluminal modes propagate at all scales.  If one is willing to allow for Lorentz violation, it is possible quarantine these 
problematic modes below some low scale with the use of multiple kinetic terms, and the modes will grow for a very short time before becoming 
super-horizon \cite{hep-th/0604153}.  And it is possible to quarantine instabilities 
to such early times that are unobservable, for example, by making the Planck mass of the auxiliary metric small in a bimetric massive gravity theory \cite{1503.07521}.  
For a low-energy effective field theory, assuming the phantom field interacts at least gravitationally, a strangely low Lorentz-violating ultraviolet cut-off of 3 MeV or below 
is needed to push instabilities to unobservable scales, 
and Lorentz-conserving cut-offs are experimentally excluded completely because of the implied modifications to gravity that would be incompatible with experiment  \cite{Cline}.

In order for a Lorentz-violating theory 
  with such a low cut-off to be viable, some low-energy unknown sector must be responsible for the ghosts, which seems improbable.  
  Perhaps one can concoct a fine-tuned theory in which low-energy effective ghosts appear 
below a very low scale while still reproducing the ghost-free standard sector below the TeV scale, but it is not clear that such a theory is feasible, especially since Lorentz-violation 
may communicate between energy sector via graviton loops \cite{hep-ph/0201082}.  
Lorentz violation is not consistent with general covariance, which is at the heart of 
the well-tested theory of general relativity.  Also, there are extremely stringent experimental 
constraints on Lorentz violation in the Standard Model \cite{hep-ph/9812418, hep-ph/9908504}.  Much work has been done on Lorentz-violating models from which the 
scientific community has learned much, and perhaps Lorentz invariance \cite{1304.5795} is broken beyond an unobservable scale that allows for the viability of such theories; this is yet to be seen.

For a theory that is Lorentz-invariant but non-local above a certain scale, it may be possible for causality to be maintained \cite{1305.3034}, and such a theory can have a 
Lorentz-invariant cut-off of $(1.8-5.6)$ meV \cite{1202.1239}, 
which is technically consistent with limits on small modifications to general relativity \cite{0611184}.  However, such a theory with non-local interactions may not satisfy 
normal $S$-matrix analyticity constraints \cite{0704.1845}.  If we observe macroscopic non-locality, 
it would be very surprising and would overturn basic assumptions we have of the physical nature of our universe.  

\section{Canonical-Sign Dark Energy and Other Frameworks with Apparent $w<-1$}

Wrong-sign field models and NEC-violating theories are clearly fraught with difficulties.  
Now we examine canonical-sign field models.  We have already shown that canonical scalar field 
theories in flat FLRW space cannot have $w<-1$, so the theory must be modified in some way to allow for at least an appearance of $w<-1$ while still having a value of $w$ 
consistent with the formulation of the canonical field theory.  And we are interested in real scalar fields because complex scalar fields, in general, imply a complex dark energy 
density, and we expect the energy density to be completely real.  

If a field or microscopic description for dark energy were not necessary, dark energy modeled as a fluid with $w<-1$ would be physically and observationally 
acceptable for a perfect fluid model, or even for more general fluid models, such as when viscosity is present \cite{1706.02543}.  However, we expect a field 
description to be fundamental, and we have already discussed the problems with a scalar field model with $w<-1$.  

 Typically, $w_\phi <-1$ is inferred from cosmological data assuming $w = p_\phi/\rho_\phi$, dark energy is a perfect fluid, and the FLRW metric.  Under these assumptions, 
 the NEC implies $P_\phi + \rho_\phi = \rho_\phi(1+w) \geq 0$ during dark energy domination, so this along with $\rho_\phi \geq 0$ 
 implies $2KE_\phi \geq 0$ as we saw earlier.  Every perfect fluid model 
 can be framed as a (non-)canonical scalar field theory, but not every scalar field theory can be framed as a perfect fluid.  Energy-momentum tensors containing 
 terms with second derivatives cannot be modeled as perfect fluids \cite{1208.4855, 1103.5360, 1106.0292}.  For 
 an imperfect fluid, or a theory that lacks perfect homogeneity and isotropy, 
 the NEC will manifest as a different inequality, and adherence to the NEC may not automatically imply positivity of the 
 kinetic energy term in the Lagrangian.  However, in general, the NEC is tied to the sub-luminality of a field theory \cite{0912.4258}.  Ê  
 
 In theory, it is possible for a model to lead to an {\it apparent} measured value of $w <-1$ under the usual assumptions of FLRW space and dark energy as a perfect fluid while 
 having positivity of the kinetic term of the actual field theory and avoiding the pathologies discussed previously.  We discuss some examples of such scenarios below.  


\subsection{Photon-Axion Conversion with Apparent $w<-1$}

Cs{\'a}ki {\it et al} \cite{astro-ph/0409596, hep-ph/0111311}, show how the magnitudes of supernovae may be dimmed by photon-axion 
conversion enough to result in an inference from the data of a rate of acceleration faster than the actual one.  They show that an inferred value for $w$ from supernovae data would 
be 
\begin{equation}
w \simeq -1 - (2.13 \Omega_m+0.04)(L_{dec}H_0)^{-1},
\label{axionw}
\end{equation}
where $L_{dec}$ is the decay length.  A cosmological constant with a sufficient rate of photon-axion conversion, consistent with all observational constraints on axions, can lead 
to a value of $w$ as low as $-1.5$.  

\subsection{Weakening Gravity in the Infrared to Achieve an Apparent $w<-1$}

Modifying gravity in the IR is another avenue that offers an apparent $w<-1$.  Carroll {\it et al} \cite{astro-ph/0408081} investigate whether scalar-tensor theories 
can result in an apparent $w<-1$ while still honoring the NEC and avoiding phantom 
status and the associated pathologies.  They examine Brans-Dicke scalar field theory, covering a broad class of scalar-tensor models, and conclude that this is possible given 
the observational constraints on the time-dependence of Newton's constant $G$.  The Friedmann equations are modified due to the the modified gravity Lagrangian, so an 
apparent measured value of 
$w<-1$ in the framework using general relativity and FLRW space can be provided from a Brans-Dicke theory that does not contain a phantom field.  
However, such a theory would need to be fairly fine-tuned.  In general, the Brans-Dicke scalar potential needs to be such that the field is near a maximum at present with a small 
first derivative with respect to time and a large second derivative. 

Another approach, by Sahni and Shtanov \cite{astro-ph/0202346}, uses a class of braneworld models in which the scalar curvature of the induced brane metric 
contributes to the brane action.  The spatially flat braneworld can exhibit acceleration while still satisfying the Randall-Sundrum constraint on the brane.  Their model leads to 
modified Friedmann equations, which allow for $w<-1$ in the usual framework of general relativity and FLRW space while suffering from no phantom pathologies.  It is even 
possible for the the late-time acceleration phase to come to an end, making the theory more amenable to the requirements of string theory.  Similarly, an apparent $w<-1$ is also 
achieved in the DGP braneworld model by Lue and Starkman \cite{astro-ph/0408246}.  

\subsection{Quintessence Field with Apparent $w<-1$}

It is possible to achieve an apparent $w<-1$ without leaving the confines of the well-vetted theory of general relativity.  
Cs{\'a}ki {\it et al} demonstrate that a quintessence field riding up a mild uphill section of its potential, after having gained kinetic energy from riding down a slope, can lead to a 
measured constant value of $w_\phi<-1$ \cite{astro-ph/0507148}.  The standard determination of the equation-of-state parameter $w_\phi$ involves the FLRW luminosity 
distance in the magnitudes of the supernovae, and fitting the data assuming a constant $w_\phi$ gives a different equation of state compared to a non-constant assumption.  They 
show that for a quintessence potential, a decrease in $w_\phi$ from -0.73 for redshift $z > 0.47$ to $w_\phi = -1$ for $z < 0.47$ with a dark energy fraction of the total 
energy density $\Omega_{DE} = 0.80$ very closely mimics dark energy with constant $w_\phi<-1$.  Such a potential providing this 
change in $w_\phi$ would not be anymore fine-tuned 
than other potentials in the literature, and the effective theory would be perfectly normal.  This change in $w_\phi$ represents an increase in the acceleration of the universe, 
a phenomenon usually associated with phantom dark energy, while maintaining $w_\phi> -1$ throughout.  

\subsection{Considering Small Deviations to FLRW Space to Achieve an Apparent $w<-1$}

We now examine attempts within the confines of general relativity that assume small deviations from the standard FLRW metric, since we know that our physical universe 
is not perfectly homogeneous throughout.  The author of this review examined a quintessence field $\phi$ in 1st-order perturbed FLRW space \cite{1507.06492}.  The goal was to 
manifest an apparent $w < -1$ in flat FLRW space while actually maintaining $w_\phi \geq -1$ in the full, perturbed metric, thus avoiding phantom pathologies.  Using 
inflation constraints on the constants of integration in the perturbations, we showed that a constant apparent $w <-1$ was not possible with $w_\phi \geq -1$.  For some 
selected models with non-constant apparent $w<-1$, we were able to show that this may be satisfied with $w_\phi \geq -1$ for certain relevant time and length scales, but not all times 
and lengths that are covered by supernovae data.  And we suspected that if we had included matter and radiation components with their density perturbations into our calculations, 
there would have been even fewer times and length scales for which $w_\phi \geq -1$. 

However, by the same token, we showed that an apparent $w \geq -1$ in FLRW space could have $w_\phi <-1$ for the full field theory in the full, perturbed FLRW metric for certain 
time and length scales, 
indicating that a naively quintessence field theory in FLRW space would actually have phantom pathologies in the 1st-order FLRW framework.  
Similarly, Onemli and Woodard \cite{gr-qc/0204065} show that a non-phantom scalar field in classical 
FLRW space (de Sitter, specifically) can violate the NEC on cosmological scales when quantum corrections via renormalization were taken into account, making it subject to 
phantom pathologies.  G{\"u}mr{\"u}k\c{c}{\"u}o\u{g}lu {\it et al} \cite{1606.00618} examine a non-phantom scalar field model with an anisotropic 
Bianchi I background and show that there are superluminal modes present in the IR range, another exhibition of phantom pathology on the large scale.  However, they show 
that these modes may be squarely identifiable with Jeans instability, a classical phenomenon, as opposed to a quantum instability.  They point out that the spatial gradient in the 
metric was key in the formation of these superluminal modes. 

Since there were hints found of an apparent $w<-1$ being provided by a well-behaved scalar field in perturbed FLRW space in the our previous work, 
perhaps the full realization is possible when quantum corrections are taken into account, treating the scalar field as a quantum field in a perturbed FLRW space.  We are 
currently working on such a treatment \cite{future}.    

Another deviation from the FLRW assumptions of isotropy and homogeneity comes from gravitational backreaction \cite{Buchert}.  The universe may have started out fairly uniform 
according to the cosmic microwave background (CMB), but we know that it is not perfectly homogeneous due to the structures that have formed and are 
continuing to form.  So if we assume an inhomogeneous space and take the spatially average of quantities in Einstein's equations, we can take into account the local effects 
on space-time expansion due to inhomogeneities in the universe.  However, it does not seem that backreaction effects can account for a present-day apparent value of 
$w<-1$ \cite{astro-ph/0612151, 0909.0749}.  A very recent constraint on the present-day apparent deceleration parameter in timescape cosmology is 
$q_0 = -0.043^{+0.004}_{-0.000}$ \cite{1706.07236}.

\section{Conclusion}

Dark energy with $w<-1$ is tenable according to cosmological data, and how this feature manifests fundamentally is still in question.  In this brief review, we have discussed 
several different attempts to ground this observation in theory.  It is mathematically inconsistent for scalar field theories to exhibit $w_\phi <-1$ in FLRW in the framework 
of general relativity.  A scalar field theory with a wrong-sign kinetic term, however, does exhibit $w_\phi <-1$.  However, this kind of theory is plagued with an unstable 
vacuum with a divergent decay rate.  The generalization to k-essence, allowing the Lagrangian density to be written as a function of the kinetic term $F(X, \phi)$, and modifications 
to general relativity are fraught with some subset of the following pathologies:  ghosts, superluminal modes, Lorentz violation, non-locality, and instability to quantum corrections.  
Perhaps a stable effective field theory that is only Lorentz-violating or non-local at certain unobservable scales, as we discussed, is physically acceptable, but this 
is not clear at the present time and seems improbable.  

The alternative to a field theory that is fundamentally phantom is one that is not phantom but still accords with a measurement of $w<-1$.  Since this observed $w<-1$ is in the 
context of general relativity with dark energy modeled as a perfect fluid, any theory that exhibits $w<-1$ in this context while remaining non-phantom in actuality is a good candidate.  
Gravity that is modified from general relativity on the cosmologically large scale is able to exhibit an apparent $w<-1$, and perhaps a field theory in a cosmological background that 
is more physically accurate than FLRW space is as well.  We plan to investigate a full quantum treatment of a dark energy field in a perturbed FLRW space in future work.  
More simply, a non-phantom scalar field theory that is rolling up its potential can also exhibit $w<-1$ observationally, 
and we discussed how the conversion of photons to axions, fully within 
observational constraints, can dim the luminosity of supernovae enough to lead to an apparent $w<-1$.  

As our cosmological probes become more and more precise, we hope to plumb more deeply the true nature of the acceleration of our universe.

\section*{Acknowledgments}

I am thankful to acknowledge support from the LaGrange College Summer Research Grant Award.



\end{document}